\documentclass{aa}
\usepackage{graphicx}
\usepackage{natbib}
\usepackage{color}
\bibpunct{(}{)}{;}{a}{}{,} 
\begin{document}

\def\ms{M$_{\odot}$}
\def\zs{Z$_{\odot}$}
\def\mga{$^{24}$Mg}
\def\mgb{$^{25}$Mg}
\def\mgc{$^{26}$Mg}
\def\msun{${\rm M_{\odot}}$}
\def\cratio{$^{12}{\rm C}/^{13}{\rm C}$}

\title{Thermohaline mixing: A physical mechanism governing the photospheric composition of low-mass giants}

\author{Corinne Charbonnel\inst{1,2} and Jean-Paul Zahn\inst{3}
        }

\authorrunning{C. Charbonnel and J.-P. Zahn}
 
\titlerunning{Thermohaline mixing in red giant stars}

\offprints{C. Charbonnel}

\institute{ Geneva Observatory, University of Geneva, chemin des Maillettes 51, 1290 Sauverny, Switzerland
\email{Corinne.Charbonnel@obs.unige.ch}
\and
Laboratoire d'Astrophysique de Toulouse et Tarbes, CNRS UMR 5572, Universit\'e Paul Sabatier Toulouse 3, 14, av.E.Belin, 31400 Toulouse, France
\and
LUTH, CNRS UMR 8102, Observatoire de Paris, 92195 Meudon, France \quad
\email{Jean-Paul.Zahn@obspm.fr}    }
\date{Accepted for publication in A\&A Letter}

\abstract{}{Numerous spectroscopic observations provide compelling evidence
for a non-canonical mixing process that modifies the surface abundances of 
Li, C and N of low-mass red giants when they reach the bump in the luminosity function.
Eggleton et al. (2006) have proposed that a molecular weight inversion created by the 
$^3$He($^3$He,2p)$^4$He reaction may be at the origin of this mixing, 
and relate it to the Rayleigh-Taylor instability.
In the present work we argue that one is actually dealing with a double 
diffusive instability referred to as thermohaline convection and
we discuss its influence on the red giant branch.}
{We compute stellar models of various initial metallicities using the 
prescription by Ulrich (1972) (extended to the case of a non-perfect gas) for the 
turbulent diffusivity produced by the thermohaline instability in stellar radiation
zones. }
{Thermohaline mixing simultaneously accounts for the observed behaviour
of the carbon isotopic ratio and of the abundances of Li, C and N on 
the upper part of the red giant branch. 
It reduces significantly the $^3$He production with respect to canonical
evolution models as required by measurements of $^3$He/H in galactic HII regions.}
{Thermohaline mixing is a fundamental physical process that must be included 
in stellar evolution modeling.}


\maketitle

\section{Introduction}

During the first dredge-up ({\sl 1dup}; Iben 1965), the surface composition 
of low-mass red giant stars\footnote{i.e., stars with initial masses below 
$\sim$~2~--~2.5M$_{\odot}$ that evolve along the Red Giant Branch (RGB) to high 
luminosities until helium is ignited in their core under degenerate conditions} 
is modified due to dilution of the external convective stellar layers 
with hydrogen-processed material: The lithium and the carbon abundances 
as well as the carbon isotopic ratio decline, while the helium 3 
and nitrogen abundances increase. The amplitude of these variations depends on 
the stellar mass and metallicity. This picture is nicely supported by observations. 

After the {\sl 1dup} the stellar convective envelope retreats (in mass) ahead
of the advancing hydrogen-burning shell (HBS) that surrounds the degenerate helium core, 
and no further surface abundance variation is predicted by canonical
stellar evolution theory\footnote{By this we refer to the modelling of 
non-rotating, non-magnetic stars, where convection is the only transport 
process considered.} on the RGB. However, spectroscopic observations clearly 
point out that some non-canonical mixing connects the convective envelope with the
HBS and further modifies the surface composition of low-mass giants as soon as 
they reach the bump in the luminosity function.
The sudden drop of the carbon isotopic ratio 
$^{12}$C/$^{13}$C provides the most pertinent clue of this mechanism 
(Gilroy 1989; Gilroy \& Brown 1991; Charbonnel 1994; Charbonnel et al. 1998; 
Gratton et al. 2000; Shetrone 2003; Recio-Blanco \& De Laverny 2007) 
that also modifies the abundances of lithium, carbon and nitrogen (e.g., 
Gratton et al. 2000). 
This non-canonical process appears to be universal as it affects at least 
95$\%$ of the low-mass stars, whether they belong to the field, to open, or 
globular clusters (Charbonnel \& Do Nascimento 1998). This high number
satisfies the galactic requirements for the evolution of the $^3$He abundance 
(Tosi 1998; Palla et al. 2000; Romano et al. 2003), 
since the mechanism which is responsible for the low values of $^{12}$C/$^{13}$C 
is also expected to lead to the destruction of $^3$He by a large factor in the bulk 
of the envelope material as initialy suggested by Rood et al. (1984; see also 
Charbonnel 1995; Hogan 1995; Weiss et al. 1996). 

At the present time, we have no firm physical model for the non-canonical 
RGB mixing. Parametrized approaches have been used to
reproduce individual observations with the diffusive velocity treated 
as a free parameter (i.e. Boothroyd \& Sackmann 1999; Weiss et al. 2000). 
On the other hand, rotation-induced mixing has been investigated thoroughly 
(see references in Charbonnel \& Palacios 2004). It turns out however 
that meridional circulation and shear turbulence alone do not produce
enough mixing to account for the surface abundance variations as
required by the observations (Palacios et al. 2006). 

\section{Mixing due to thermohaline convection}
Recently Eggleton et al. (2006, 2007) identified a possible cause for such non-canonical mixing, namely the molecular weight inversion created by the 
$^3$He($^3$He,2p)$^4$He reaction in the external wing of the HBS. As it was pointed out already by Ulrich (1972), this nuclear reaction is somewhat singular in that it produces more particles per unit mass than it started from. 
Using their 3D hydrodynamic code to model a low-mass star at the RGB tip (Dearborn et al. 2006),  Eggleton and 
collaborators found that such a $\mu$-profile leads to efficient mixing, of the kind required to ``reconcile Big Bang and stellar nucleosynthesis 
(as far as $^3$He is concerned)", as they put it.

According to them, the instability responsible for that mixing is the well-known Rayleigh-Taylor instability, a dynamical instability which is triggered when a layer of heavier fluid lies over lighter fluid. In stellar interiors, that instability takes the form of the convective instability, which tends to render the temperature gradient adiabatic rather than to suppress the density inversion.
But what first occurs in a star, as the inverse $\mu$-gradient builds up, is actually a double diffusive instability, which was discussed long ago in the literature under the generic name of ``thermohaline convection" (Stern 1960). 
This instability appears in various astrophysical situations, for instance when $^4$He or C-rich material is deposited at the 
surface of a star in a mass transferring binary (Stothers \& Simon 1969; Stancliffe et al. 2007). Recently it has been invoked when a star 
accretes heavy elements during planet formation (Vauclair 2004).

The thermohaline instability differs from the convective instability in that it involves two components, of which one, the stabilizing one (temperature) diffuses faster than the other (salt), whose stratification is unstable.
It occurs in a stable stratification that satisfies the Ledoux criterion for convective stability:
\begin{equation}
\nabla_{\rm ad} - \nabla + \left({\varphi \over \delta}\right)\nabla_\mu > 0 ,
\end{equation}
but where the molecular weight decreases with depth: 
\begin{equation}
\nabla_\mu  := {{\rm d} \ln \mu \over {\rm d} \ln P} < 0 .
\end{equation}
We use the classical notations  for $\nabla = (\partial \ln T / \partial \ln P)$, $\varphi = (\partial \ln \rho / \partial \ln \mu)_{P,T}$ and $\delta=-(\partial \ln \rho / \partial \ln T)_{P,\mu}$.

In the laboratory, the instability takes the form of ``salt fingers"; since heat diffuses faster than salt, these 
fingers sink because they grow increasingly  heavier than their environment, until they become turbulent and dissolve. In stellar interiors, the role of salt is played by a heavier species, such as helium, in a hydrogen-rich medium.

Ulrich (1972) was the first to derive a prescription for the turbulent diffusivity produced by that instability in stellar radiation zones. Through a linear analysis, and assuming perfect gas ($\varphi=\delta=1$) he gets (his eqs. 24 and 31)
\begin{equation}
D_t =  C_t \, K {- \nabla_\mu \over (\nabla_{\rm ad} - \nabla)} \quad \hbox{for} \;  \nabla_\mu <0 ;
\label{coeff-u}
\end{equation}
$K$ is the thermal diffusivity.
His non-dimensional coefficient involves the aspect ratio $\alpha$ (length/width) of the fingers:
\begin{equation}
C_t = {8 \over 3} \pi^2 \alpha^2 ;
\end{equation}
for the value he  advocates, $\alpha = 5$, this coefficient is rather large: $C_t = 658$.

Kippenhahn et al. (1980) extended Ulrich's expression to the case of a non-perfect gas (including radiation pressure, degeneracy):
\begin{equation}
D_t =  C_t \,  K  \left({\varphi \over \delta}\right){- \nabla_\mu \over (\nabla_{\rm ad} - \nabla)} \quad \hbox{for} \;  \nabla_\mu < 0 .
\label{dt}
\end{equation}
However instead of fingers they considered ``blobs" which according to them are destroyed very rapidly: their 
mixing length is of the order of the blob size and this translates into a smaller value of 12 for the coefficient $C_t$.

It is of course disturbing that 
these coefficients $C_t$ differ by almost two orders of magnitude, depending on whom one believes; 
it would be most desirable to confront them with realistic 3D simulations. In the meanwhile, we tend to favour
a large value for the coefficient $C_t$ as advocated by Ulrich, because all experiments so far have displayed slender 
fingers, rather than blobs (cf. Krishnamurti 2003). Crude as it may be, this prescription has the advantage of being 
rooted in the physical process, unlike the admittedly ad hoc diffusivity chosen by Eggleton et al. (2007).
 
\section{Model calculations}
We compute three evolution models of a 0.9~M$_{\odot}$ star with various 
initial values of [Fe/H], namely --1.8, --1.3 and --0.5, with the aim
of comparing our theoretical predictions with observations over a large
range of metallicity. 
We use the code STAREVOL (Siess et al. 2000; Palacios et al. 2003, 2006). 
In the present study the transport of particles in the radiative regions is due 
to thermohaline mixing only. The expression we use for $D_t$ is that given in Eq.~\ref{dt}
that includes the correction for a non-perfect gas.
For simplicity we assume $C_t = 1000$, which is of the order of magnitude of 
the coefficient advocated by Ulrich (1972). The model at [Fe/H]=--1.3 is also 
computed with $C_t = 100$ and 12. We do not consider any additional transport 
process related to rotation. 

\begin{figure}
\centering
\resizebox{\hsize}{!}{\includegraphics{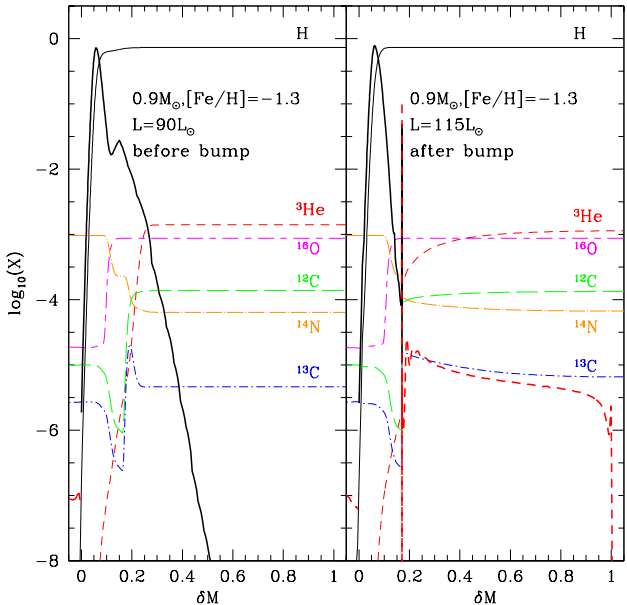}}
\caption{Profiles of the logarithm of the mass fraction of H, $^3$He, $^{12}{\rm C}$,
  $^{13}{\rm C}$, $^{14}{\rm N}$, $^{16}{\rm O}$, and of 
  the mean molecular weight gradient $\nabla_\mu$ = d $\ln \mu$/ d $\ln P$ (bold line, 
  full when $\nabla_\mu > 0$, dashed otherwise) 
  as a function of the reduced mass (see text) inside a 0.9 \msun, [Fe/H]=--1.3 star. 
  The left and right panels correspond to models including thermohaline mixing 
  and located just before and after the bump. 
  In the latter case, the thermohaline instability has just set in naturally, while
  in the former case it is still prevented due to the strong $\mu$-gradient left 
  behind by the {\sl 1dup}
}
\label{profilsabond}
\end{figure}

Fig.~\ref{profilsabond} presents the abundance profiles of selected elements 
and the mean molecular weight gradient $\nabla_\mu$ = d~$\ln \mu$/ d$\,\ln P$, just before and after 
the bump in the 0.9 \msun, [Fe/H]=--1.3 star when $C_t$=1000. The abscissa is the 
relative mass $\delta M=(M_r - M_{\rm core})/(M_{\rm env} - M_{\rm core})$ 
defined as ranging from 0 to 1 between the bottom of 
the HBS and the base of the convective envelope. 

During the {\sl 1dup} phase, the convective envelope homogenizes the star
down to very deep regions, and builds a very steep gradient of molecular weight at
the point of its maximum penetration. This corresponds to the external peak 
(at $\delta M \sim 0.12$) in the profile of $\nabla_\mu$ that can be seen 
in the left panel of Fig.~\ref{profilsabond}. 
On the other hand the deeper peak (at $\delta M \sim 0.05$) locates 
the region where H is efficiently depleted by nuclear reactions in the HBS. 
One sees that at the current luminositity (L=90L$_{\odot}$) both peaks are very close 
to each other, meaning that the star is approaching the bump. At this evolutionary 
point, $\nabla_\mu$ is positive in the whole radiative region, and the thermohaline 
instability cannot set in (see Eq. 2). The profiles of the chemical elements 
are thus identical to those obtained in a canonical model. 

When the HBS passes through the $\mu$-discontinuity left behind by the {\sl 1dup} 
(i.e., at the bump), H-burning occurs in a homogeneous region. 
The $^3$He($^3$He,2p)$^4$He reaction slightly lowers the molecular weight in 
the external wing of the HBS where $\nabla_\mu$ becomes negative (bold dashed 
line in the right panel of Fig.~\ref{profilsabond}), allowing the
thermohaline instability to develop naturally between the $^3$He-burning region 
and the base of the convective envelope. Deeper inside the radiative region, 
$\nabla_\mu$ remains positive (bold full line) and no thermohaline mixing occurs.
As can be seen in Fig.~\ref{profilsabond}, the surface abundance of $^3$He, 
$^{12}$C, $^{13}$C, and $^{14}$N are already modified soon after the onset
of thermohaline mixing. However the surface abundance of $^{16}$O remains 
constant because the thermohaline mixing does not extend down to the very deep 
region where full CNO-burning operates at equilibrium. 

\begin{figure}
\centering
\resizebox{0.8\hsize}{!}{\includegraphics{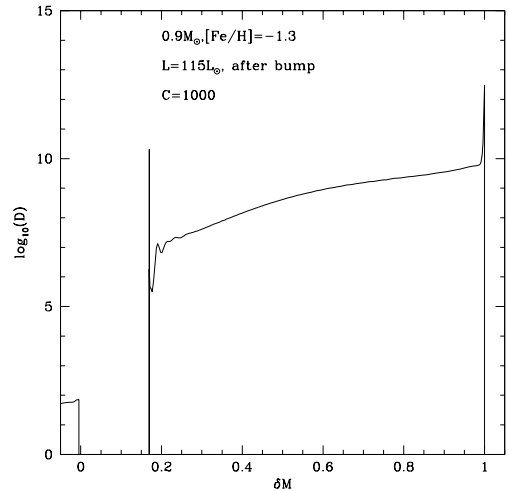}}
\caption{Logarithm of the diffusion coefficient $D_t$ (assuming $C_t$=1000) inside the 
  0.9 \msun, [Fe/H]~=~--1.3 star just after the bump.
}
\label{profilsDthermohaline}
\end{figure}

We show in Fig.~\ref{profilsDthermohaline} the profile of the diffusion coefficient 
$D_t$ (Eq. 5, for $C_t$=1000), for the same model just after the onset of the thermohaline
instability (L=115L$_{\odot}$). $D_t$ is high enough to connect the HBS and the 
convective envelope and to modify the surface abundances as shown below. 
It is two to three orders of magnitude higher than that characterizing the rotational
mixing (Palacios et al. 2006).

\section{Signatures of mixing and comparison with observations}

\begin{figure}
\centering
\resizebox{\hsize}{!}{\includegraphics{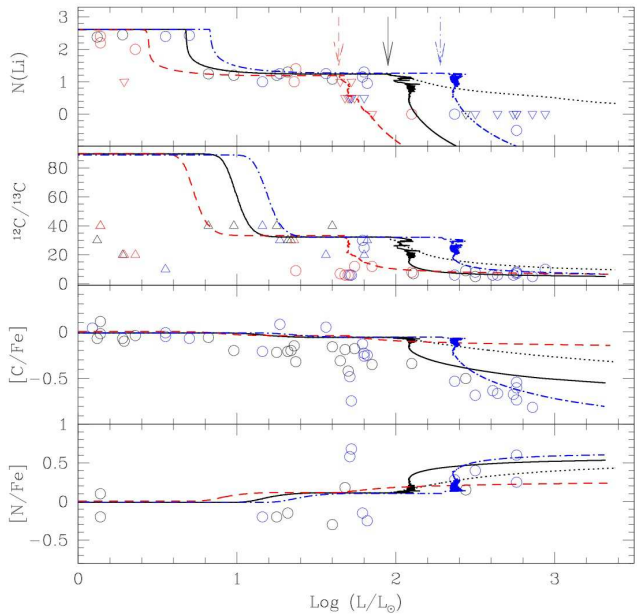}}
\caption{Evolution of the lithium abundance, of  the carbon isotopic 
  ratio \cratio, of [C/Fe] and [N/Fe] as a function of the luminosity logarithm 
  for the 0.9M$_{\odot}$ models with [Fe/H]=--1.8, --1.3 and -0.5 (respectively 
  dashed-dotted blue, full line black, and dashed red) computed with 
  $C_t$=1000 in Eq. 5. 
  The black dotted line is for the model with [Fe/H]=--1.3 calculated with $C_t$=100. 
  The arrows in the upper panel indicate the location of the bump for the three metallicities.
  Observational data are from Gratton et al. (2000) for field stars 
  in the metallicity range [Fe/H] $\in$ [-2;-1]. 
  Blue symbols are for stars with measured [Fe/H] values lower than --1.4,
  black for stars with $-1.4\leq$[Fe/H]$\leq-1.2$, 
  and red for [Fe/H] higher than --1.2. 
  Circles are actual measurements, open upward triangles are {\em lower}
  limits and open downward triangles are {\em upper} limits. 
}
\label{figcompgratton00}
\end{figure}

In Fig.~\ref{figcompgratton00} we compare the temporal evolution of 
$^7$Li, $^{12}$C/$^{13}$C, [C/Fe], and [N/Fe], obtained at the surface 
of our models with homogeneous observational data for field stars
with --2$\leq$[Fe/H]$\leq$--1 by Gratton et al. (2000). 
We see first that variations associated with the {\sl 1dup} fit well 
the observational behaviour on the lower RGB. The surface abundances 
then stay constant until the stars reach the bump. Until that evolutionary point 
the predictions are identical to those of the canonical models. 

Then at the luminosity of the bump (which increases for a given 
stellar mass when the initial metallicity decreases, as shown by the arrows) 
thermohaline mixing leads to a second episode of abundance variations that 
explains remarkably well the data\footnote{Oscillations in stellar luminosity can be seen 
just above the bump. They are related to the transport of $^3$He from the convective
envelope into the outer HBS region where this element burns in favour of $^7$Be. 
The related nuclear energy release is large. When it happens to rival that of the pp-chains, 
the total luminosity increases and the star readjusts. Once the abundance of $^3$He is low enough
the oscillations vanish.}
Li is rapidly destroyed, and the
carbon isotopic ratio reaches values between 5 and 7 very close from the equilibrium 
value. [C/Fe] and [N/Fe] respectively decrease and increase. These 
C and N variations are stronger for lower initial metallicity, 
as required by the observations. 
As already noted in \S~3, the 
surface abundances of the heavier elements ($^{16}$O and $^{23}$Na in
particular, not shown here) which are affected by nuclear reactions 
much closer to the He-core do not vary at the surface. 

The models shown in Fig.~\ref{figcompgratton00} were computed assuming
$C_t=1000$ in Eq. 5. We also show the predictions for the model with [Fe/H]=--1.3
computed with $C_t=100$. In that case the surface abundances change less 
rapidly and one does not reproduce the abrupt variation at the luminosity
of the bump depicted by the observations. Our analysis is thus compatible with 
the high value for the coefficient $C_t$ advocated by Ulrich (1972) and it
excludes the low value (12, not shown here) derived by Kippenhahn et al. (1980).

\section{The case of $^{3}$He}
\begin{figure}
\centering
\resizebox{0.8\hsize}{!}{\includegraphics{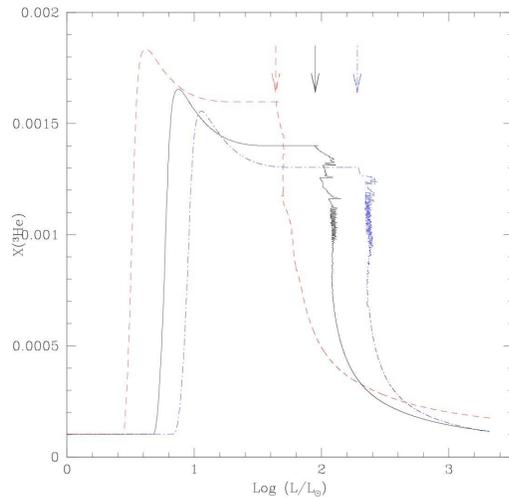}}
\caption{Evolution of the surface abundance of $^3$He (in mass fraction).
 The line symbols are as in Fig.~\ref{figcompgratton00}.
}
\label{3He}
\end{figure}

On the main sequence, a $^{3}$He peak builds up due to pp-reactions 
inside the low-mass stars (Iben 1967), and is engulfed in the stellar envelope
during the {\sl 1dup}. As a consequence the surface abundance of $^3$He
strongly increases on the lower RGB as can be seen in Fig.~\ref{3He} 
for our models at three different metalicities.
Its value reaches a maximum when the whole peak is engulfed. 
After the {\sl 1dup} the temperature at the base of the convective envelope
is too low for $^{3}$He to be nuclearly processed. As a result in 
canonical models $^{3}$He stays constant at the surface and its final 
value is strongly increased with respect to the initial one (this is the 
value before thermohaline mixing sets in at the bump). 

After the bump, thermohaline mixing brings $^{3}$He from the convective
envelope down to the region where it is nuclearly burned. This leads 
to a rapid decreases of the surface abundance of this element as 
can be seen in Fig.~\ref{3He}. 
This confirms the early suggestion by Rood et al. (1984) that 
the variations of the carbon isotopic ratio and of $^{3}$He are strongly
connected (see also Charbonnel 1995; Eggleton et al. 2007). 
It is important to note that in the models presented here $^{3}$He decreases by a large factor 
in the ejected material with respect to the canonical evolution predictions
but that low-mass stars remain net producers (while far much less efficient 
than in the canonical case) of $^{3}$He.
Computations for different masses and metallicities have now to be performed 
in order to estimate the actual contribution of low-mass stars to 
Galactic $^{3}$He in the framework proposed here. 
We are confident that the corresponding $^{3}$He yields will 
help reconciling the primordial nucleosynthesis with measurements 
of $^{3}$He/H in galactic HII regions (Balser et al. 1994; Bania et al. 2002;
Charbonnel 2002). 

\section{Conclusions}
It is known since a long time that a non-canonical mixing process 
modifies the surface abundances of low-mass stars when they arrive at 
the bump in the luminosity funtion. 
Based on 3D-modeling of a low-mass star at the RGB tip, Eggleton and 
collaborators have proposed that the molecular inversion created 
by the $^3$He($^3$He,2p)$^4$He reaction in the external wing of the HBS
may be the root cause for such a mixing. They ascribe this mixing
to the Rayleigh-Taylor instability.

In the present work we argue that one is actually dealing with a double
diffusive instability called thermohaline convection,  which has been
discussed long ago in the literature. 
We compute stellar models including the prescription given by Ulrich (1972) 
and extended to the case of a non-perfect gas 
for the turbulent diffusivity produced by that instability 
in stellar radiative zone. The results presented here indicate that 
thermohaline convection simultaneously accounts for the observed behaviour 
of the carbon isotopic ratio and for the abundances of Li, C and N in RGB stars
at and above the bump in the luminosity function.
It also avoids large $^3$He production by low-mass stars as required
by chemical evolution models of the Galaxy.   
It does not modify the O nor the Na surface abundances. 
In a future work the effect of the thermohaline mixing in models of various masses 
will be presented. 

\begin{acknowledgements}We thank the referee for her/his comments that helped us improve 
our manuscript and Dr Eggleton for sending us his paper before publication.
C.C. is supported by the Swiss National Science Foundation (FNS). 
We acknowledge travel support from the French Programme National de Physique Stellaire (PNPS). 
\end{acknowledgements}

{}

\end{document}